\documentclass{raa}            % referee version: for submission

\usepackage{xcolor}
\definecolor{RedAdd}{RGB}{200,0,0}    % 新增内容:深红色（RGB值对应期刊常用色）
\definecolor{BlueDel}{RGB}{0,0,200}   % 删除内容:深蓝色（避免打印后看不清）
\usepackage{graphicx,times}             %for PS/EPS graphics inclusion, new
\usepackage{natbib}
\usepackage{amssymb,amsmath}
\bibpunct{(}{)}{;}{a}{}{,}

\usepackage[pagebackref=true]{hyperref}

\begin{document}

  \title{Estimation of  gravitational wave from solar emerging  magnetic flux tube
}
  % \subtitle{GW from sun}

   \volnopage{Vol.0 (20xx) No.0, 000--000}      %%preserved for Editor. DOn't remove!
   \setcounter{page}{1}          %%starting page, preserved for Editor. DOn't remove!

   \author{Siqi Guan %(周爱英) %% Put your Chinese name in "( )" if you like. Note to open line 11 "\usepackage[UTF8]{ctex}"
      \inst{1,2*}
   \and Shangbin Yang
      \inst{1, 2*}
   \and Xiao Guo
      \inst{3*}
   }
%% Here is an example of three authors come from different institutes.
%% For single author or all the authors from an institute, use "\inst{}" only

   \institute{State Key Laboratory of Solar Activity and Space Weather, National Astronomical Observatories, Chinese Academy of Sciences,
             Beijing 100101, China \\
%% Please give the E-mail address of the author, to whom future correspondence and
%% offprint requests will be sent.
 \and
             School of Astronomy and Space Science, University of Chinese Academy of Sciences, 19A Yuquan Road, Beijing 100049, China\\
        \and
             School of Fundamental Physics and Mathematical Sciences, Hangzhou Institute for Advanced Study, University of Chinese Academy of Sciences, Hangzhou 310024, China\\
\vs\no
   {\small *Correspondence: guansiqi22@mails.ucas.ac.cn(G.S.); yangshb@nao.cas.cn (Y.S.) ; guoxiao17@mails.ucas.ac.cn (G.X.)}}

    \abstract{This study investigates the gravitational waves (GWs) generated by the emergence of magnetic flux tubes in the solar convection zone. We focus on the upward buoyancy of magnetic flux tubes, which leads to significant magnetic activity and the formation of active region sunspots. This study adopts parameters representative of a moderate-sized solar active region to estimate the GWs generated by the emergence of magnetic flux tubes. Our results indicate that the GW strain amplitude, achievable through signal superposition and detection at close proximity (e.g., approximately one solar radius from the solar surface), may reach $\sim$10$^{-29}$. The characteristic GW frequency is estimated at $\sim$10$^{-5}$ Hz, placing it at the high-frequency end of the sensitivity band of Pulsar Timing Array (PTA) methods. However, the estimated strain amplitudes remain orders of magnitude below the sensitivity thresholds of current and foreseeable gravitational wave detectors. Notably, reducing the cadence $\Delta t$ of Pulsar Timing Array (PTA) observations to approximately 2 hours ($\Delta t = 2\text{hours}$) would raise the maximum detectable frequency to about $5.8 \times 10^{-5} \text{Hz}$, thereby encompassing the dominant spectral component of solar activity-related GWs predicted in this study, offering a potential pathway for future detection. Successful detection in the future may help to predict the super solar active region emergence in space weather forecasting.
}

   \authorrunning{Guan, Yang \& Guo}            %author_head in even pages
   \titlerunning{Estimation of  gravitational wave from solar emerging  magnetic flux tube}  % title_head in odd pages

   \maketitle
%% The author head (on even pages) and the title head (on odd pages) will be
%% automatically extracted from \author{} and \title{}. Whenever the title is too long,
%% you will be asked to supply a shorter one by inserting either \authorrunning{} or
%% \titlerunning{} before \maketitle. Anyway, you can specify your own heads.
%%
%%
%% Note: In the following text body of your manuscript, please note several differences from
%%       other major journals:
%% (1) \subsection{Please Capitalize the First Letter of Each Notional Word in Subsection Title}
%% (2) Please Capitalize the First Letter of Each Notional Word in all tables' captions

%
%________________________________________________ sections below
%
\section{Introduction}

The solar convection zone governs the Sun's magnetic activity, significantly influencing space weather and stellar physics. \citet{gizon2005local} challenge traditional buoyant flux tube models by demonstrating how toroidal fields rise to form active regions, underscoring the need to probe magnetic dynamics in this critical layer. This research emphasizes the importance of studying the solar structure and the role of magnetic fields in solar activity, highlighting the need for continued investigation into the solar convection zone's magnetic dynamics. To probe these complex dynamics deep within the Sun, researchers have historically relied on two primary methods: the historical exploration of the Sun's interior has primarily utilized helioseismology and solar neutrino observations. Helioseismology, through the analysis of solar oscillations, has been instrumental in mapping the Sun's interior structure
  \citep{fan2021magnetic}. Solar neutrinos, detected through various experiments, have provided direct evidence of the nuclear reactions occurring within the Sun's core \citep{bahcall1989neutrino}. These methods have significantly contributed to our understanding of the internal dynamics of the Sun.

  The quest to understand the formation of sunspots on the Sun has been a significant endeavor in solar physics. Historically, scientists have been attempting to predict the appearance of sunspots by tracking the rise of magnetic fields from the Sun's interior \citep{babcock1961topology}. Despite these efforts, accurately forecasting sunspot emergence has proven elusive. In a breakthrough,  \citep{ilonidis2011detection} have detected magnetic fields forming deep within the Sun by helioseismology, approximately 60, 000 kilometers beneath the surface, just one to two days before sunspots appear. Researchers detected a marked increase in the emergence rate of magnetic flux, which can cause a noticeable acceleration in the travel time of sound waves which serves as an indicator of an impending sunspot emergence. While this study offers a promising lead for space weather prediction and enhances our grasp of solar magnetic field dynamics, the precise timing and intensity of solar eruptions associated with sunspots still remain an open question.

Gravitational waves (GWs), as predicted by Einstein's theory of general relativity, were first detected in 2015 by the LIGO Scientific Collaboration and the Virgo Collaboration, marking a significant milestone in physics \citep{Abbott2016}. This direct observation confirmed a key prediction of general relativity and launched a new era of astronomical observation. GWs, unlike electromagnetic radiation, do not experience significant scattering or absorption as they propagate through the solar interior, making them an attractive tool for probing the internal structure of the Sun. This unique property allows for the potential detection of previously inaccessible regions within the solar convection zone and tachocline. GWs, unimpeded by solar plasma, offer a novel probe: solar magnetic flux emergence generates GWs via quadrupole moments, which could provide a new avenue for fundamental research into solar dynamics. \citet{kokkotas1999quasi} pointed out that the detection of GWs produced by solar oscillations or magnetic activity, if detected, could provide valuable information about the Sun's interior structure and dynamics. The potential of GWs for solar interior probing has been highlighted in several theoretical frameworks \citep{hanasoge2012anomalously}. The propagation of GWs in the solar medium has been extensively studied, with recent advancements in numerical simulations providing insights into the behavior of these waves within the context of solar dynamics \citep{brdar2019gravitational}.   \citet{2023ApJ...957...52T}       proposes leveraging lensed GWs from pulsars to probe solar density, bypassing limitations of intrinsic solar GW detection and traditional methods. \citet{GarciaCely:2024sun} calculate the complete GW spectrum generated by solar interior plasma, encompassing both microscopic (particle collisions) and macroscopic (hydrodynamic fluctuations) mechanisms. It assesses their detectability and contribution to the high-frequency GW background. These studies suggest that GWs could offer a novel perspective on the deep interior of the Sun, complementing traditional helioseismic methods.

Note that asymmetric magnetic flux emergence produces a non-zero gravitational quadrupole moment, generating detectable GWs. The Sun's proximity offers a unique detection opportunity, yet quantitative GW estimates from emerging flux tubes are scarce—particularly for predicting super active regions vital to space weather \citep{Wang2009}.  This study aims to fill this theoretical gap by providing a first-order model and estimate, which is valuable for understanding the energetics of solar active regions. In this study, we model flux tubes as catenary curves—a geometry motivated by force equilibrium
bu using observational parameters of emerging solar  active regions in solar cycle 23-24. We estimate the intensity of GWs generated during magnetic flux emergence and explore the potential for detection by GW observatories. In Sec.\ref{sec:mod}, we introduce the single-source oscillation model, adopting the catenary model to simplify the shape of magnetic flux tubes. Through quasi-static and magnetic flux conservation assumptions, we derive an equivalent density. The results are then discretized and incorporated into the single-source oscillation GW model, ultimately obtaining the amplitude-frequency relationship of GWs. In Sec.\ref{sec:dis}, we discuss the detectability of etimated GWs by detectors at different locations based on the estimated GW strength, and also consider the detectability after signal superposition.

\section{MODELING AND RESULTS}
\label{sec:mod}

\subsection{Single-source GW model}
\label{ss gw}

Within the framework of general relativity, the generation of GW requires the second order derivative of the mass quadrupole moment $\ddot{Q}_{ij}$ must be non-zero. The emergence process of magnetic flux tubes in the solar interior perfectly satisfies these conditions—the upward motion of flux tubes from the convection zone (approximately $0.75R_\odot$) to the photosphere exhibits significant spatial asymmetry, while the acceleration driven by magnetic buoyancy ($d^2r/dt^2 \sim B_\phi^2/(4\pi\rho r^2)$) produces strongly time-varying quadrupole moments. Moreover, radial density gradients ($\nabla\rho$) in the plasma further amplify quadrupole variations. The quadrupole moment holds central importance in GW detection: the wave amplitude $h \propto |\ddot{Q}_{ij}|/(c^4 r)$ depends directly on the quadrupole change rate, representing the dominant non-zero multipole order for gravitational radiation and carrying crucial information about the source's internal dynamics.

The theoretical foundation of GW generation lies in the quadrupole formula, which relates wave amplitude to the second time derivative of the mass quadrupole moment.  For GWs propagating through spacetime, the transverse-traceless (TT) gauge amplitude is expressed as:

\begin{equation}
h_{ij}^{\rm TT}(t,\mathbf{x}) = \frac{1}{r}\frac{2G}{c^{4}} \Lambda_{ij,kl}(\hat{\mathbf{n}}) \ddot{Q}_{kl}(t-r/c)
\label{eq:quad_amp}
\end{equation}
in which $r$ denotes the source-observer distance; $G$ and $c$ represent the gravitational constant and speed of light respectively; $\Lambda_{ij,kl}$ is the projection tensor enforcing transverse-traceless conditions; $\ddot{Q}_{kl}$ signifies the second time derivative of the reduced quadrupole moment:

\begin{equation}
Q^{ij} = \int \rho(\mathbf{x}) \left( x^i x^j - \frac{1}{3}r^2\delta^{ij} \right) \mathrm{d}^3x
\label{eq:quad_moment}
\end{equation}
This fundamental relationship connects the dynamics of mass-energy distributions with spacetime curvature perturbations.

The gravitational wave polarization amplitudes \(h_+\) and \(h_\times\) are derived from the second time derivative of the reduced quadrupole moment \(\ddot{Q}^{ij}\) (defined in Eq.\ref{eq:quad_moment}). However, in practical calculations, it is often convenient to work with the second mass moment \(M^{ij}\):
\[
M^{ij} = \int \rho(\mathbf{x})  x^i x^j  d^3x
\]
The reduced quadrupole moment \(Q^{ij}\) is related to \(M_{ij}\) by:
\[
Q^{ij} = M^{ij} - \frac{1}{3} \delta^{ij} M^k_k
\]
where \(M^k_k = \text{Tr}(M)\) is the trace of the mass moment tensor. For GW generation, only the trace-free part of \(M^{ij}\) contributes to radiation, which is exactly \(Q_{ij}\).

GWs manifest through two polarization modes, $h_+$ and $h_\times$, whose amplitudes depend on the wave propagation direction relative to the source orientation. For propagation along the $z$-axis ($\hat{\mathbf{n}} = \hat{\mathbf{z}}$), the polarization components simplify to:

\begin{equation}
h_{+} = \frac{1}{r} \frac{G}{c^4} \left( \ddot{M}_{11} - \ddot{M}_{22} \right)
\end{equation}
\begin{equation}
 h_{\times} = \frac{2}{r} \frac{G}{c^4} \ddot{M}_{12}
\end{equation}

For arbitrary propagation directions $\hat{\mathbf{n}} = (\sin\theta\cos\phi, \sin\theta\sin\phi, \cos\theta)$, the complete angular dependence is described by:

\begin{align}
h_{+}(t;\theta,\phi) &= \frac{G}{c^{4}r} \Big[
  \ddot{M}_{11}(\cos^2\phi - \sin^2\phi\cos^2\theta) + \ddot{M}_{22}(\sin^2\phi - \cos^2\phi\cos^2\theta) \nonumber \\
  &- \ddot{M}_{33}\sin^2\theta - \ddot{M}_{12}\sin 2\phi(1 + \cos^2\theta) \nonumber \\
  &+ \ddot{M}_{13}\sin\phi\sin 2\theta + \ddot{M}_{23}\cos\phi\sin 2\theta \Big] \\
h_{\times}(t;\theta,\phi) &= \frac{G}{c^{4}r} \Big[
  (\ddot{M}_{11} - \ddot{M}_{22})\sin 2\phi\cos\theta + 2\ddot{M}_{12}\cos 2\phi\cos\theta \nonumber \\
  &- 2\ddot{M}_{13}\cos\phi\sin\theta + 2\ddot{M}_{23}\sin\phi\sin\theta \Big]
\label{eq:pol_general}
\end{align}

These expressions fully characterize the wave's angular dependence and polarization states for arbitrary source configurations.

Following \citet{hanasoge2008}'s formulation of convective elements as vertically oscillating mass sources, we similarly model emerging magnetic flux tubes as point masses undergoing acceleration along the radial direction. This approach reduces complex magnetohydrodynamic processes to an analytically tractable quadrupole moment $\ddot{Q}_{zz}$ (in Hanasoge's notation) or $\ddot{Q}_{33}$ (in our notation), enabling efficient amplitude estimation.
To model GWs from emerging solar magnetic flux tubes, we approximate a flux tube element as a point mass oscillating along the $y$-axis. The density distribution is described by:

\begin{equation}
\rho(t,\mathbf{x}) = \mu(t)  \delta(x)\delta(z)\delta(y - y_0(t))
\label{eq:point_mass}
\end{equation}

where $\mu(t)$ represents the effective mass and $y_0(t)$ its time-dependent position. This simplification preserves essential dynamics while rendering the quadrupole moment tractable. The resulting mass moment is:

\begin{equation}
M^{ij} = \mu(t)  y_0^2(t)  \delta^{i3}\delta^{j3}
\label{eq:mass_moment}
\end{equation}

Substitution into polarization equations produces axisymmetric waveforms:

\begin{align}
h_{+} &= -\frac{1}{r}\frac{G}{c^{4}} \ddot{M}_{33} \sin^2\theta \\
h_{\times} &= 0
\label{eq:final_waveform}
\end{align}

\begin{figure}[ht]
 \centering
  \includegraphics[width=0.8\linewidth]{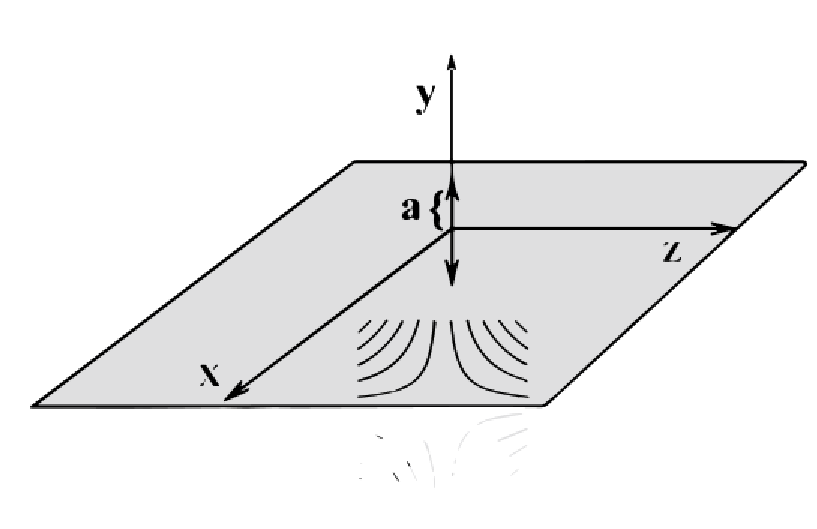}
  \caption{GW Model for Single-Source: The source oscillating along the $y$ axis (double arrow), and the transverse stretching - compressing mode of the GW in the direction of $\theta=\frac{\pi}{2}$(schematic of field lines), a represents the amplitude. Source: \citep{Maggiore08}.
  }
  \label{fig:single}
\end{figure}

The amplitude scaling $h \sim \ddot{M}_{33}/r$ reveals two critical aspects for detection: 1) the requirement for strong accelerations in dense plasma regions, and 2) the advantage of proximity to the source. These insights guide our subsequent analysis of solar GW detectability. Fig. \ref{fig:single} illustrates the physical configuration and resulting wave pattern. The axisymmetric nature originates from cylindrical symmetry in the quadrupole moment generated by motion along a single axis. This simplified model establishes the foundation for estimating GWs from solar magnetic flux emergence, where complex dynamics are reduced to an effective vertical oscillation.

\begin{figure}[ht]
 \centering
\includegraphics[width=0.8\linewidth]{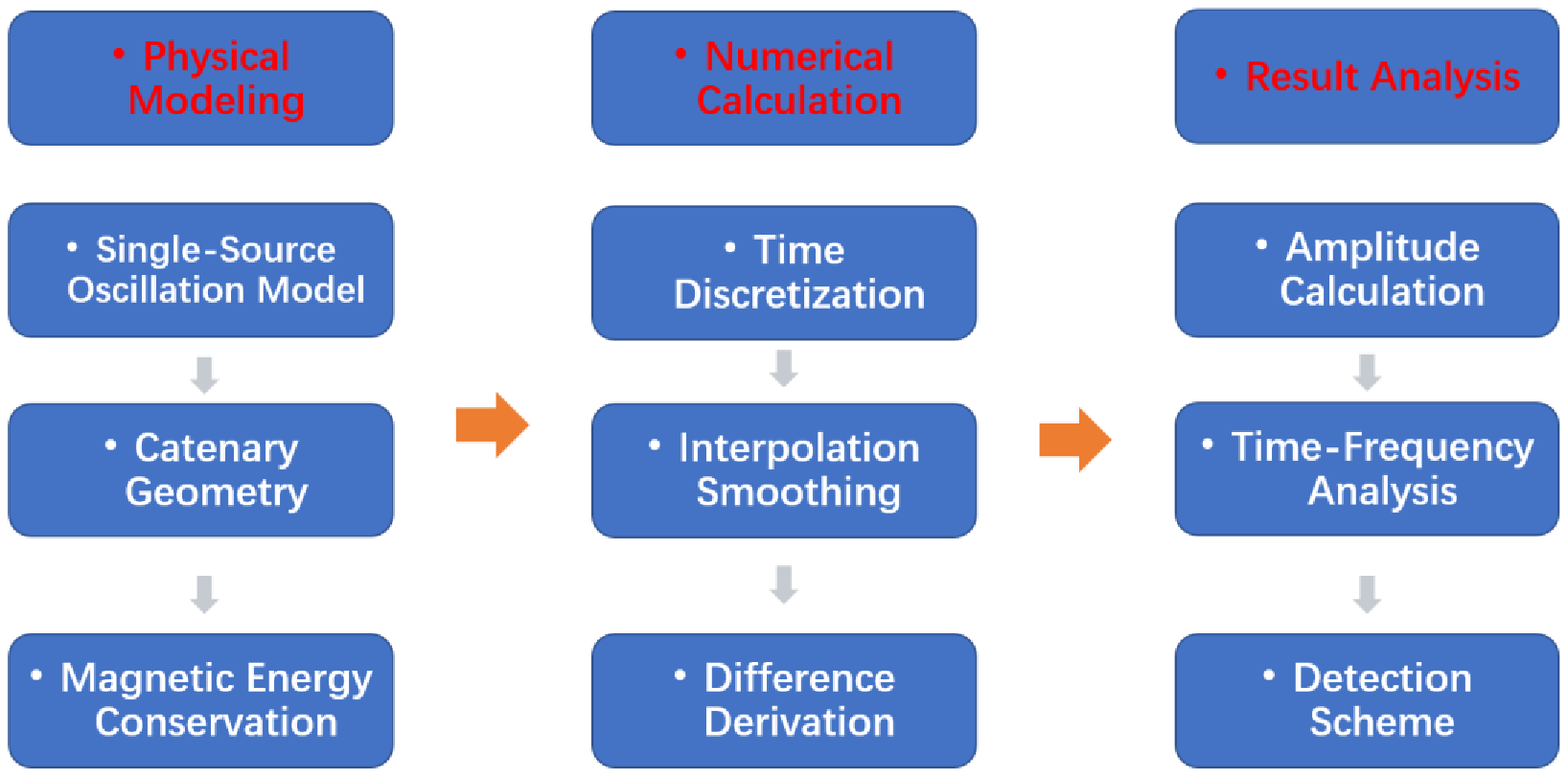}
  \caption{Flowchart of the gravitational wave estimation framework. }
  \label{fig:flow_chart}
\end{figure}

After introducing the single-source GW model in this study,
fig.\ref{fig:flow_chart} illustrates the logical flow of our estimation approach. Firstly, we build the physical model, in which the upward movement of solar magnetic flux tubes is simplified as a cluster of catenary curves with fixed endpoints. The geometric parameters are determined by the intersection coordinates of the catenary equation with the solar surface, and the equivalent density at different heights is derived using the principle of conservation of magnetic energy. Secondly, we present the numerical calculation process. The ascent of the emerging magnetic flux tube is discretized into several time steps, assuming the center point undergoes uniformly accelerated linear motion to obtain the displacement-time relationship. Finally, the result and analysis are presented. According to the single-source oscillation model used in this study, cubic spline interpolation is used to smooth the data, and the central difference method is applied to calculate the second derivative of the mass quadrupole moment, thereby obtaining the spatiotemporal distribution of GW amplitude.
This whole above process transforms complex magnetohydrodynamic processes into a computable quadrupole radiation model through catenary geometry, magnetic energy conservation, and motion discretization.

In the following subsection, Sec.\ref{2.2} shows that how we adopt a catenary curve model to capture the quasi-static ascent of magnetic flux tubes. Sec.\ref{2.3} is to derive the equivalent density profile  by following the geometric setup. Sec.\ref{2.4} and Sec.\ref{2.5} describe how to apply the motion discretize and compute the quadrupole moment. Sec.\ref{2.6} shows the results of estimated GW amplitude and the time-frequency spectrograh.

\subsection{Catenary model of Flux Geometry}
\label{2.2}

The adoption of the catenary geometry and the subsequent simplification of the flux tube motion to a uniformly accelerated ascent rely on two key physical assumptions: the quasi-static nature of the rise and the dominance of magnetic forces leading to nearly constant acceleration. These approximations are justified within the context of solar interior conditions and are supported by established magnetohydrodynamic (MHD) principles.

The rise of magnetic flux tubes through the solar convective zone is characterized by an extremely high magnetic Reynolds number (\(R_m \sim 10^{11}\)), a dimensionless parameter that quantifies the dominance of magnetic convection over diffusion. This condition indicates that the magnetic field is perfectly ``frozen'' into the highly conductive plasma on the emergence timescales considered in this study, satisfying Alfvén's frozen-in theorem. This physical regime justifies the treatment of the ascent within the framework of a quasi-static assumption.
The core of this approximation lies in the clear separation of timescales. The characteristic rise time (\(\tau_{\text{rise}} \sim \text{days}\)) is significantly longer than the Alfvén time (\(\tau_A = L / v_A \sim \text{hours}\)), the timescale required for the flux tube to establish mechanical equilibrium internally and with its surroundings:
\begin{equation}
\tau_{\text{rise}} \gg \tau_A.
\end{equation}
This disparity implies that the flux tube evolves through a series of near-equilibrium states, rendering the inertial term (\(\rho D\mathbf{v}/Dt\)) in the momentum equation negligible. Consequently, the dynamics are governed by a sequential static force balance at each point in the ascent:
\begin{equation}
0 \approx -\nabla p + \frac{1}{4\pi} (\nabla \times \mathbf{B}) \times \mathbf{B} + \rho \mathbf{g}_{\text{eff}}.
\label{eq:force_balance}
\end{equation}
This system of equations, coupled with the induction equation under the frozen-in flux condition, defines the quasi-static rise path of the flux tube. It is this foundational equilibrium that validates the use of a static catenary geometry to model the flux tube's shape at each instance during its ascent in our analysis.
The catenary geometry provides a mathematically tractable framework for flux tube ascent. In this study, the process of solar magnetic flux emergence is simplified to the emergence of a single catenary from the solar convection zone (at a depth of 0.75$R_\odot$) to the solar surface (at the solar radius). The analysis is confined to a two-dimensional scenario, considering only the $x$ and $y$ axes, to examine the behavior of a single magnetic flux tube within a solar active region over a period of 30 days to emerge \citep{Fan1993}.
\begin{figure}[ht]
 \centering
  \includegraphics[width=0.8\linewidth]{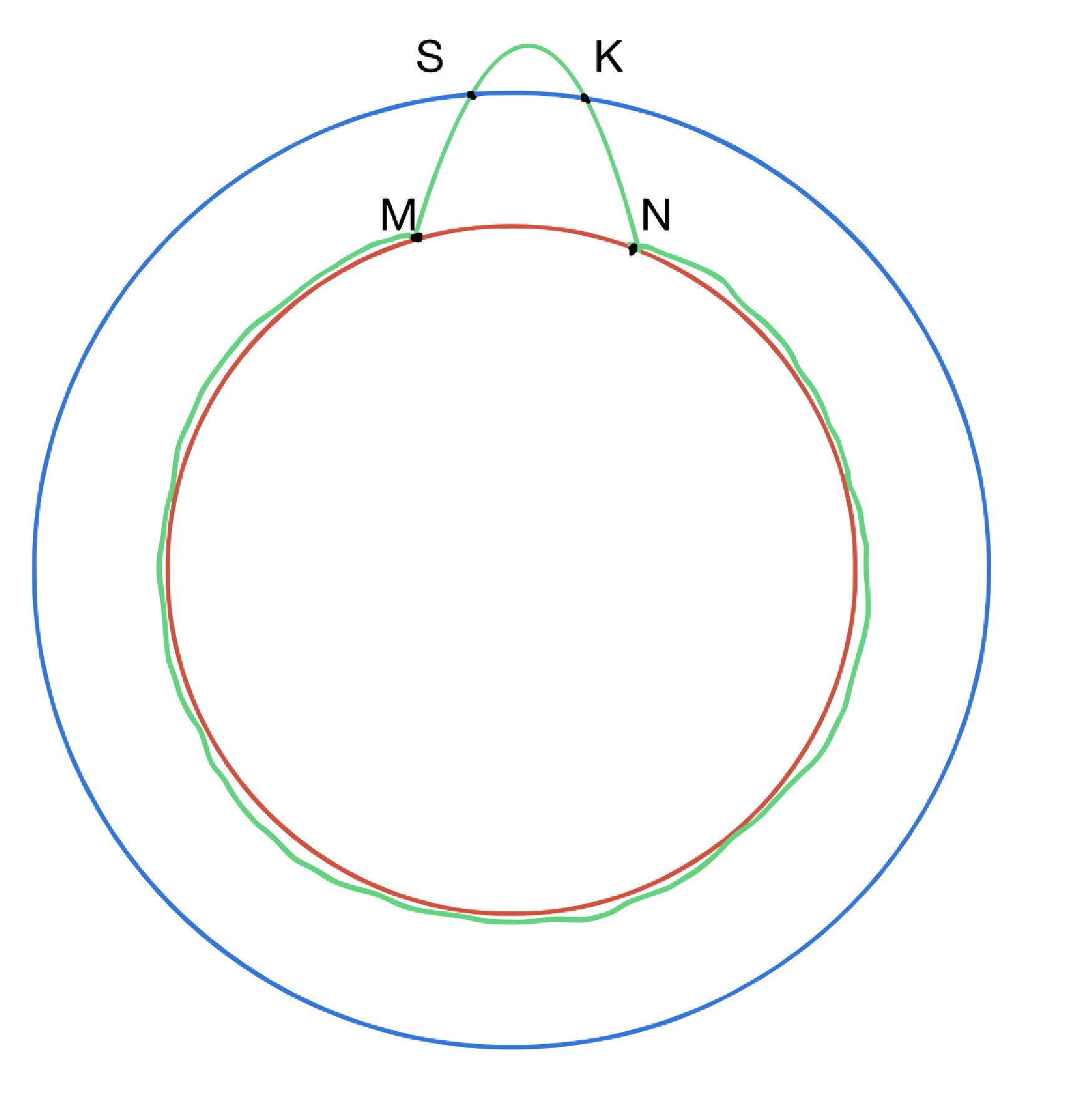}
  \caption{Catenary model, red line represents bottom of convection zone, green line represents the most marginal catenary,  blue line represents the solar surface. }
  \label{fig:model}
\end{figure}
First of all, in this paper, the emerging magnetic flux tubes in two-dimensional coordinates are simplified as a cluster of infinitely many catenary curves as Fig.\ref{fig:cluster}. The core of the catenary model lies in the exact mathematical correspondence between its differential equation form and the equilibrium equations of magnetic flux tubes. The classical catenary equation describes the static equilibrium of a flexible chain in a uniform gravitational field:
\begin{equation}
    \frac{d^2 y}{dx^2} = \frac{1}{a} \sqrt{1 + \left( \frac{dy}{dx} \right)^2}
    \label{eq:catenary_classic}
\end{equation}
in which $a$ represents the ratio of tension to linear density.
\begin{figure}[ht]
 \centering
  \includegraphics[width=0.8\linewidth]{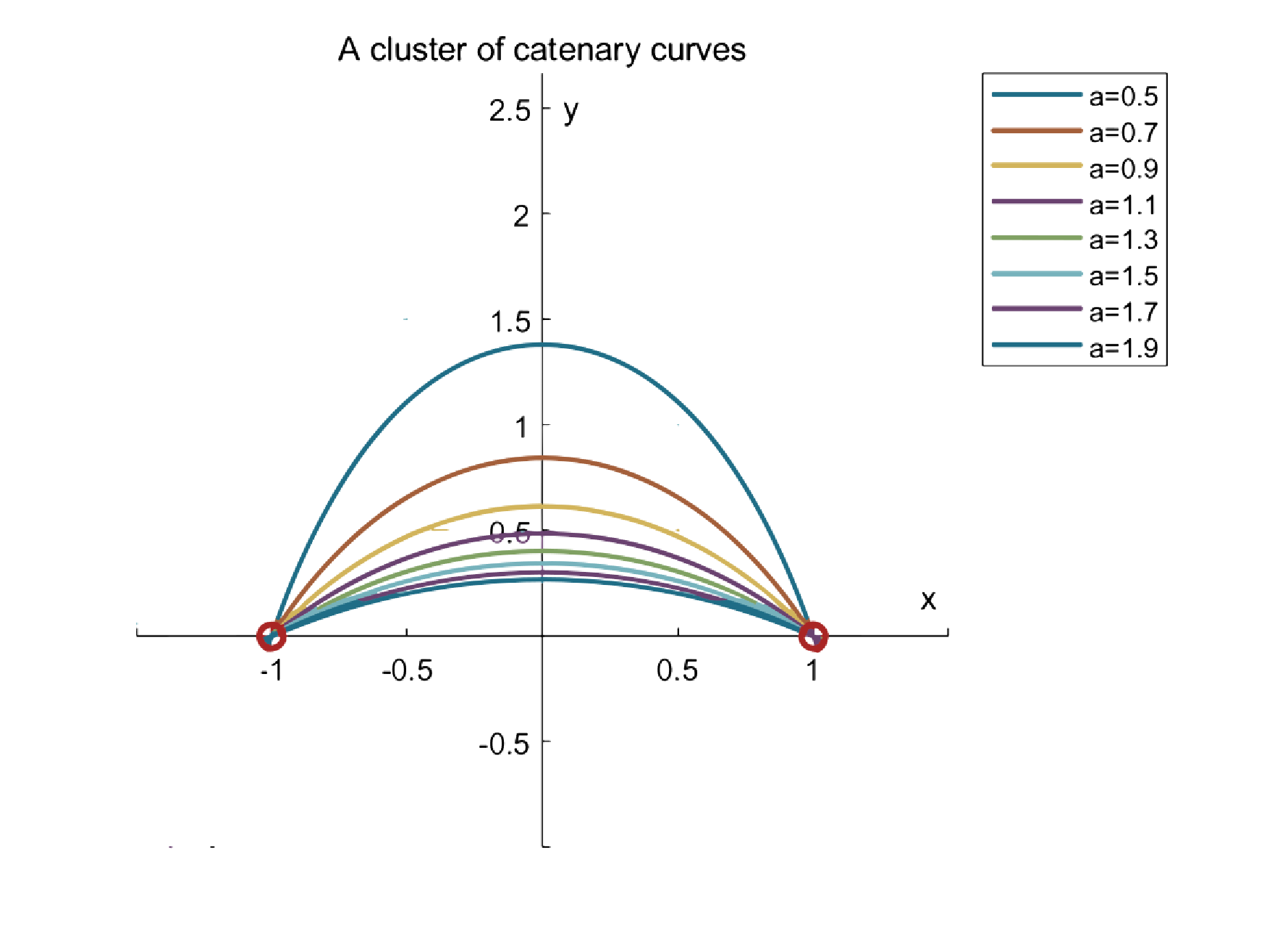}
  \caption{The figure illustrates a cluster of downward-opening catenary curves, sharing coincident x-axis intersections, as an approximation to the two-dimensional magnetic flux tube model. }
  \label{fig:cluster}
\end{figure}
The catenary model of magnetic flux tube in the sun is shown in Fig.\ref{fig:model}. The process of magnetic flux emergence from the solar convection zone to the photosphere is approximated as the vertical rise of a single catenary from the bottom of solar convective zone to the solar surface.  The vertex of the catenary, where it intersects the y-axis, is determined based on the average height of sunspots above the photosphere, which is estimated to be at a position $(0, 1.125R_\odot)$, where $R_\odot$ is the solar radius. The standard catenary equation involves two unknowns:
\begin{equation}
    y=-{a(t)}\left[\frac{e^\frac{x}{{a(t)}}+e^\frac{-x}{{a(t)}}}{2}\right]+{{b(t)}}
\end{equation}
 thus necessitating an additional condition for its full determination.

This condition is provided by the arc length between two points, (0, $R$) and K: ($p$,  $q$), on the catenary. The arc length is inferred from the radius of the active region and is approximated to be about $AL_0$ \citep{Antiochos1982SolarAR}.
\begin{equation}
    \int_0^p\sqrt{R^2-x^2} dx=AL_0
\end{equation}
With these parameters, we can derive the abscissa of K, then we substitute it into the equation of the solar surface, can we get the exact point K. Then the equation of the catenary at the vertex can be resolved, providing a mathematical description of the magnetic flux tube's shape as it emerges into the solar atmosphere.

\subsection{Density Derivation}
\label{2.3}

Under the above catenary geometry setup,  we then need to estimate
the density in the magnetic flux tube to further calculate the density
derivation. Firstly, we put forward the assumption of energy conservation, and its verification is as follows:
The magnetic Reynolds number $R_m$, as a dimensionless parameter characterizing the coupling strength between the magnetic field and fluid motion, is defined as the ratio of magnetic convection effect to diffusion effect: $\frac{UL}{\eta}$,
where $U\sim10^3$ m/s represents the typical convective velocity; $L\sim 10^8$ m represents the characteristic length of the magnetic flux tube (comparable to the active region scale); $\eta =\frac{1}{\mu_0\sigma}\sim 1$ m$^2$/s represents the magnetic diffusivity in the solar convective zone.
In the solar convective zone ($R\approx 0.7R_\odot$), $R_m \sim 10^{11}$, and the magnetic field is "frozen" into the plasma, satisfying the Alfvén's frozen-in theorem:
\begin{equation}
    \frac{\partial \mathbf{B}}{\partial t} = \nabla \times (\mathbf{v} \times \mathbf{B})
\end{equation}
In this case, the magnetic flux through any material surface is conserved.
Although the classical diffusivity $\eta_{turb}$ is negligible, turbulent motions in the solar convective zone may introduce an effective magnetic diffusion effect. Using mixing-length theory to estimate the turbulent diffusivity: $\eta_{turb}\approx0.1v_{turb}l_{turb}\sim 10^{-3}$ m$^2$/s, and the corresponding magnetic flux decay time is much longer than the flux tube rise time. This order-of-magnitude difference confirms that in single flux tube emergence events, the destruction of magnetic flux conservation by turbulent diffusion can be regarded as a higher-order small quantity.
If magnetic reconnection events are considered, when the magnetic field gradient $\nabla B>0.1$ T/m, current sheet formation leads to local reconnection with a magnetic flux loss rate of approximately 5\%. In this paper, this point is neglected, and it is still considered that the magnetic flux is approximately conserved during the rise of magnetic flux tubes in the solar interior. By invoking the assumption of conservation of magnetic energy in the vertical direction, we acknowledge that the magnetic energy present in the photosphere is:
\begin{equation}
    E^0_s=S_{\rm spots}\cdot 2L^H_{\rm pho}\cdot \frac{B^2_{\rm pho}}{2\mu _0}
\end{equation}
in which $S_{\rm spots}$ is the area of a hemisphere of a sunspot activity region, which can also be written as $\pi p^2$ (in which, $p$ represents the radius of the sunspot. Here we adopt 5000 km as the sunspot radius for moderately active regions.), $\mu _0=4 \pi \times 10^{-7}N/A^2$, $L^H_{\rm pho}$ is the semidiameter of the catenary at the solar photospheric level, and $\frac{B^2_{\rm pho}}{2\mu _0}$is the magnetic energy density within the photosphere. Secondly, we acknowledge that the magnetic energy present in any time is:
\begin{equation}
    E^0_s=2L^H(t)\cdot \pi \omega ^2_d\cdot \frac{B^2_E}{2\mu _0}
\end{equation}
in which $\pi\omega^2_d$ is the area defined by the radii corresponding to the $x$-coordinates of the intersection points of each catenary with a circle, $L^H(t)$is the semidiameter of the catenary at any given time $t$, and $\frac{B^2_E}{2\mu_0}$ is the magnetic energy density of the corresponding time.
 Solving the two aforementioned magnetic energy equations simultaneously yields:
\begin{equation}
     B^2_E=\frac{p^2\cdot 580\pi\cdot B^2_{\rm pho}}{L^H(t)\cdot \omega_d^2}
\end{equation}

According to mass energy equation: $E=mc^2=\rho \cdot V\cdot c^2=V\cdot \frac{B^2_E}{2\mu_0}$, we can yield :
\begin{equation}
    \rho=\frac{B^2_E}{2\mu_0c^2}
\end{equation}
Let's substitute $B^2_E$   into the equation above:
\begin{equation}
    \rho =\frac{p^2\cdot 580\pi \cdot B^2_{\rm pho}}{{2L^H(t)}\cdot \omega_d^2\cdot \mu _0 c^2}
\end{equation}

\subsection{Motion Discretization}
\label{2.4}

In this section, we discretize the entire emerging motion cycle of the magnetic flux tubes. To simplify the calculations, we further consider the rise of the magnetic flux tube as a uniform accelerated straight-line motion, thereby obtaining the displacement-time relationship \( s(t) \) for the center point of the magnetic flux tube. A time interval of 1000 seconds was selected to discretize the total duration of 30 days: $k=3600\times24\times 30=2.592\times 10^6 \rm  s$, resulting in 2592 distinct time steps. Consequently, there are 2592 instances of the catenary equation that need to be solved. As illustrated in  Fig.\ref{fig:model}, all 2592 catenaries share the same endpoints, denoted as M and N, with the caveat that points M and N are symmetric with respect to the $y$-axis, thus constituting a single boundary condition. It is essential to determine the coordinates of the intersection of each catenary with the $y$-axis.

Referring to the literature by  \citep{fan2021magnetic}, the motion of magnetic flux tubes is governed by the magnetohydrodynamic (MHD) equations, which account for the combined effects of magnetic fields, plasma, and gravity. Assuming radial motion of the flux tube and axisymmetry ($\frac{\partial}{\partial\theta}=0$), the equation simplifies to:
\begin{equation}
    \rho \frac{d^2 r}{dt^2} = -\frac{\partial p}{\partial r} + \frac{B_\phi}{4\pi r} \frac{\partial (r B_\phi)}{\partial r} + \rho g_{\text{eff}} - \frac{1}{r} \frac{\partial}{\partial r} (r \Pi_{rr})
\end{equation}
Here, $B_\phi$ is the azimuthal magnetic field component, $\rho $ is the plasma density, $r$ is the radial position of the flux tube. This system incorporates contributions from the internal pressure gradient, magnetic field effects, and external effective gravity. Under magnetically dominant conditions, the pressure gradient and viscous terms become negligible, reducing the equation to: $\frac{d^2 r}{dt^2} \approx \frac{B_\phi^2}{4\pi \rho r^2}$. If the azimuthal magnetic field \(B_\phi\) varies slowly with radius (\(\frac{\partial B_\phi}{\partial r} \approx 0\)), the acceleration \(a\) approaches a constant value, supporting the uniform acceleration hypothesis. Then the ascent of the catenary's central point is simplified to uniformly accelerated linear motion. The continuous motion equation of the catenary's midpoint can be derived from the duration of motion and the distance traveled as follows

\begin{equation}
    s(t)=\frac{R}{2k^2}\left[k^2e^\frac{k-t}{k}-0.5t^2\right]+\frac{R}{2k}\cdot et+(0.75-0.5e)R.
\end{equation}
Our approach involves further discretizing this motion into 2592 discrete points. This discretization also provides the second condition necessary for ascertaining the catenary equations. Thus we can discretize the catenary that rises along the $y$ axis as Eq.\eqref{eq:point_mass}
in which $y_0(t)$ is the 2592 interval time of $s(t)$. Besides, it is obvious that in the process of the magnetic flux tube rising, only the "half - stroke" of the motion along the positive direction of the $y$ axis is considered in Fig.\ref{fig:single}.

\begin{figure}[ht]
  \centering
  \includegraphics[width=0.8\textwidth]{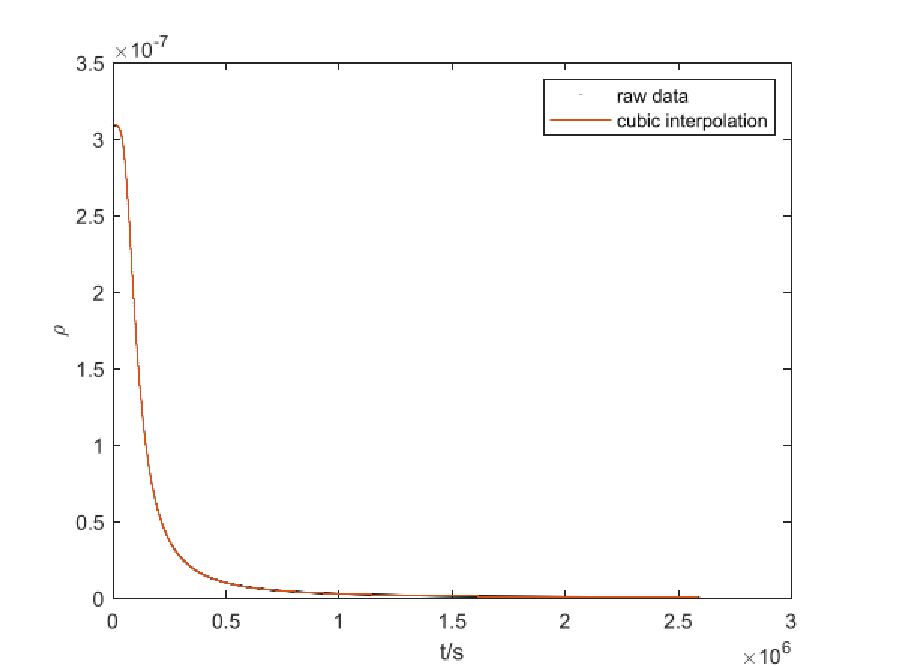}
  \caption{Temporal evolution of the equivalent density $\rho$ during flux tube ascent. The orange curve is the cubic interpolation fitting curve based on the original data, and the 2591 black dots represent raw data points.}
  \label{fig:rho}
\end{figure}

Finally, We can get $\rho $  of 2592 time intervals based on Eq.~(8).  For constructing a cubic polynomial function to approximate the unknown function between the data points, ensuring that the interpolation results are smooth and the derivatives are continuous in the adjacent intervals, which is achieved by using the cubic spline interpolation method. Fig.~\ref{fig:rho} shows the temporal evolution of the equivalent density $\rho$ during flux tube ascent.

Subsequently, we use the model in Sec.~\ref{ss gw} to calculate the GW quadrupole moment for a single source oscillating along the $y$-axis in three-dimensional space \citep{Maggiore08}.

\subsection{Quadrupole moment compuation}
\label{2.5}

\begin{figure}[ht]
  \centering
  \includegraphics[width=0.8\linewidth]{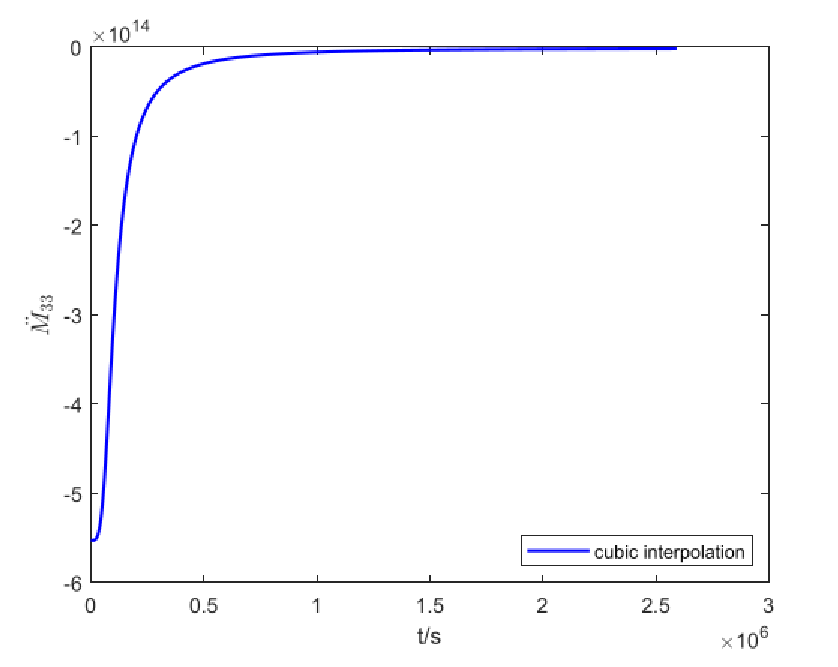}
  \caption{Temporal evolution of the second derivative of mass moment  $\ddot{M}^{33}$ . The blue curve is the cubic interpolation fitting curve based on the original data.}
  \label{fig:M33}
\end{figure}

The second mass moment is:
\begin{equation}
\begin{aligned}
    M^{ij}&=\int \rho (t, x)x^ix^j\, d^3x\\
    &=\int \rho (t, x)x^ix^j\, d^3x\\
    &=\mu (t)y_0^2(t)\delta ^{i3}\delta ^{j3}\\
    &=\mu (y)s^2(t)\delta ^{i3}\delta ^{j3}
 \end{aligned}
\end{equation}
Based on the symmetry relations, i.e., the geometric relation, it's obvious that there is only $M^{33}$ left: $M^{33}=\mu _0(t)y_0^2(t)=\mu (y)\cdot y^2_0(t)=\rho (y)\cdot y^2_0(t)\cdot \frac{H_{\rm tot}}{2592}\cdot \frac{W_{\rm tot}}{2592}$
in which, $H_{\rm tot}$ represents for the total height of the ascending process: \textbf{$(1.125-0.75)R$}, $W_{\rm tot}$ represents for the total width of the ascending  process.
\begin{equation}
\left\{
\begin{aligned}
    M^{33}&= \rho (t)\cdot y^2_0(t)\cdot \frac{(1.125-0.75)R}{2592}\cdot \frac{2m}{1000}, \\
    y_0(t) &= \frac{R}{2k^2}[k^2e^{\frac{k\cdot t}{k}}-\frac{1}{2}t^2]+\frac{R}{2k^2}\cdot ke\cdot t +(\frac{3}{4}-\frac{e}{2})R, \\
    k&= 30\times 3600\times 24.
\end{aligned}
\right.
\label{S-equations}
\end{equation}
In numerical analysis, the second-order difference is used to approximate the second derivative of a function. It is an important tool in the discretization of differential equations and signal processing. Given a set of discrete points with equal step sizes $f_i=f(x_i)$, where $x_i=x_0+ih$, the Taylor expansions of $f(x_{i+1}))$ and $f(x_{i-1})$ are as follows:
\begin{equation}
    \begin{aligned}
     f(x_{i+1})=f(x_i)+hf'(x_i)+\frac{h^2}{2}f''(x_i)+\frac{h^3}{6}f'''(x_i)+O(h^4)\\
     f(x_{i-1})=f(x_i)-hf'(x_i)+\frac{h^2}{2}f''(x_i)-\frac{h^3}{6}f'''(x_i)+O(h^4)
    \end{aligned}
    \label{Discrete Difference}
\end{equation}
Add the two equations in equation to eliminate the first-order derivative terms.
\begin{equation}
    f(x_{i+1})+f(x_{i-1})=2f(x_i)+h^2f''(x_i)+O(h^4)
    \label{After Discrete Difference}
\end{equation}
We can then obtain $f''(x_i)$. That is, using the 2592 sample points $M_{33}$ obtained from Eq.~\eqref{S-equations}, we substitute them into Eq.~\eqref{After Discrete Difference} and use the central difference method to obtain 2591 values of the second derivative of mass moment $\ddot{M}_{33}$.  Fig.~\ref{fig:M33} shows the temporal evolution of  $\ddot{M}^{33}$.

In addition, Table 1 presents the symbol used and its interpretations in this section.

 \begin{table}[ht]
\centering
\caption{symbol interpretation table in the paper}
\begin{tabular}{|c|p{12cm}|}
\hline
symbol & explanation \\ \hline
$R_\odot$ & solar radius \\ \hline
$k$ & total duration of single emergence cycle \\ \hline
$t$ & time \\ \hline
$s(t)$ & displacement-time relationship of the center point \\ \hline
$e$ & nature exponential \\ \hline
$S_{spots}$ & area of a hemisphere of a sunspot activity region \\ \hline
$E^0_s$ & magnetic energy present in the photosphere \\ \hline
$L^H_{pho}$ & semi-diameter of the catenary at the photosphere \\ \hline
$\mu_0$ & permeability of vacuum \\ \hline
$B_{pho}$ & magnetic field intensity of photosphere \\ \hline
$L^H(t)$ & semi-diameter of the catenary of time t \\ \hline
$\omega_d$ & radii corresponding to x-coordinates of intersection points of each catenary with a circle \\ \hline
$B_E$ & magnetic field density of corresponding time \\ \hline
$c$ & speed of light \\ \hline
$\rho$ & density \\ \hline
$m$ & mass \\ \hline
$M^{ij}$ & second mass moment \\ \hline
$H_{tot}$ & total height of ascending process \\ \hline
$W_{tot}$ & total width of ascending process \\ \hline
$h_+$ & polarization component of + mode \\ \hline
$h_{\times}$ & polarization component of cross mode \\ \hline
$AL_0$ & arc length \\ \hline
\end{tabular}
\vspace*{-0.5cm}
\end{table}
\label{Table:parameters}
\clearpage

\begin{figure}[ht]
  \centering
  \includegraphics[width=0.8\linewidth]{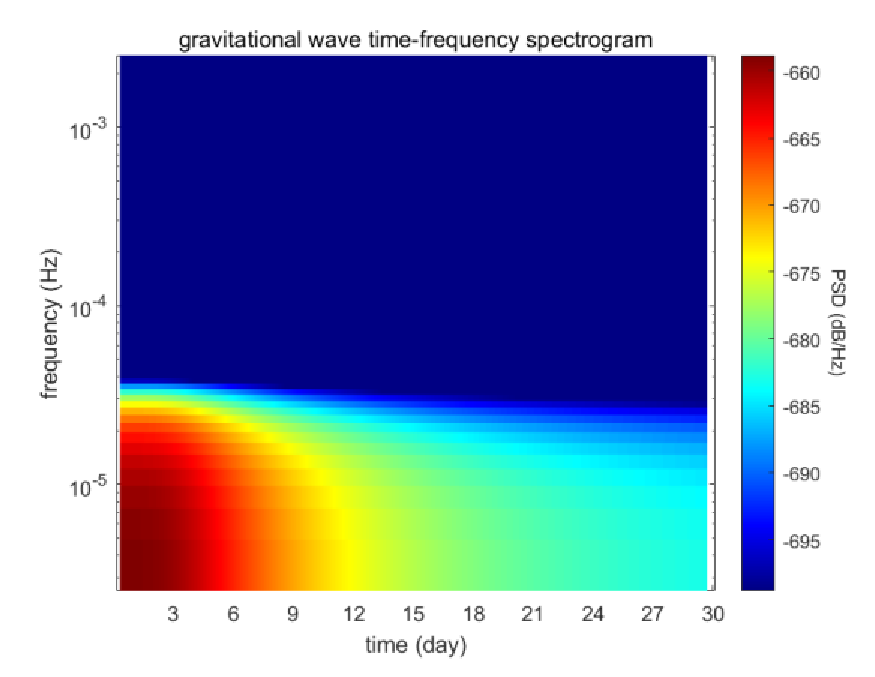}
  \caption{The time-frequency spectrogram obtained through SFFT method, in which colors are used to represent the logarithmic power spectral density of the GW strain signal across time and frequency, illustrating the distribution of signal intensity.}
  \label{fig:t_f} \vspace*{-0.3cm}
\end{figure}

\subsection{RESULTS}
\label{2.6}

Once $\ddot{M}^{33}$ is given in last subsection, we are easily able to compute the angular distribution of the quadrupole radiation:
\begin{equation}
    h_+(t; \theta, \phi)=-\frac{1}{r}\frac{G}{c^4}\ddot{M}^{33(tret)}\sin^2{\theta}\\
    =\frac{2G\mu a^2\omega^2_s}{rc^4}\sin^2\theta \cos(2\omega_s t_{ret})
    \end{equation}
    \label{equ h}
\begin{equation}
h_\times(t; \theta, \phi)=0
\end{equation}
Here, $r$ denotes the distance between the detector and the wave source,
$G$ is the gravitational constant, $G = 6.67430 \times 10^{-11}\rm m^{3}kg^{-1}s^{-2}$.

Based on GW strain signal Eq.\eqref{equ h}, we perform a Short-Time Fourier Transform (STFT) to obtain the time-frequency spectrogram in Fig.\ref{fig:t_f}. It intuitively displays the dynamic evolution of frequency components in GW signals over time. As shown in Fig.\ref{fig:t_f}, the frequency range is lower than $3\times 10^{-5}\rm Hz$, which is close to the PTA band\citep{bustamante2022gravitational}.

By integrating these profiles into our theoretical framework and assuming a detector at
\begin{equation}
  r = {1}{\rm AU}
\end{equation}
we derive a GW strain amplitude of
\begin{equation}
  h \sim 10^{-42}
  \label{eq:strain}
\end{equation}
This value is far below the detectability range of existing GW observatories.Notably, the characteristic frequency of gravitational waves (GWs) generated by solar emerging magnetic flux tubes—estimated herein to be ~\(10^{-5}\ \text{Hz}\) (as the dominant component in Fig.7)—exhibits striking consistency with the sensitive frequency band of the Cassini spacecraft, a landmark low-frequency GW detector. According to  \cite{Armstrong2006}, Cassini conducted systematic GW observations via precision Doppler tracking during its 2001–2003 solar opposition campaigns, covering a frequency range of \(10^{-6}–10^{-3}\ \text{Hz}\) . This frequency alignment is not coincidental. Cassini’s \(10^{-6}–10^{-3}\ \text{Hz}\) band was strategically designed to detect low-frequency GWs from astrophysical sources with spatial scales comparable to the Earth-spacecraft separation (~1–10 AU), which aligns with the physical scale of solar magnetic flux tube emergence (from solar radius to AU-scale). Importantly, this consistency does not imply that Cassini directly detected GWs from solar flux tubes; rather, it confirms that our theoretically derived frequency falls within the validated range of mature low-frequency GW detection technologies, providing indirect observational support for the plausibility of our model—where GW generation is driven by the buoyancy of magnetic flux tubes.The superposition of signals from multiple flux tubes could potentially approach the detectability threshold for future high-precision GW observatories.

%In summary of this section, we focus on the GW emission resulting from the emergence of a single magnetic flux rope in a solar eruption. The temporal evolution of the equivalent density \( \rho \), as shown in Fig.\ref{fig:cluster},  reveals a dynamic phase transition during the flux rope ascent, captured by cubic interpolation (orange curve) of key observational data points (diamonds). This density evolution directly drives the quadrupole moment dynamics, with the second-order difference of the mass moment \( \ddot{M}_{33} \) exhibiting a sharp peak in Fig.\ref{fig:rho}, consistent with the impulsive nature of magnetic reconnection.

\section{DISCUSSION}
\label{sec:dis}

To properly contextualize the significance of our estimated signal, it is essential to compare it not only in amplitude but also in character to the dominant sources of the stochastic gravitational wave background (GWB) in the nanohertz to microhertz regime. The GWB in this frequency band is expected to be dominated by: (i) a cosmological component from processes such as inflation or phase transitions in the early universe, and (ii) an astrophysical component from the unresolved population of supermassive black hole binaries (SMBHBs) \citep{Agazie2023, Sesana2016}.

The key distinction lies in their temporal properties. Both the cosmological and SMBHB backgrounds are modeled as stationary, Gaussian, and isotropic random processes. Their timescales are cosmological or galactic, spanning from years to the age of the universe ($\sim 10^7 - 10^{17}$ seconds), making them effectively constant over any realistic observational cadence \citep{Maggiore2018, LISA2017}.

In stark contrast, the GW signal from solar magnetic flux emergence, as predicted by our model, is intrinsically transient and non-stationary. It is tied to discrete emergence events with a characteristic duration on the order of days. Crucially, its dominant oscillatory component has a period of $\sim$1 day ($\sim 10^5$ seconds), as determined by the buoyant rise time.

This fundamental difference in timescales—cosmological ($\sim 10^{17}$ s) / stellar-orbit ($\sim 10^7$ s) vs. solar-dynamic ($\sim 10^5$ s)—is the key to distinguishing a potential solar contribution. It implies that the solar signal would be localized in time and manifest as a non-Gaussian, intermittent component superimposed on the smooth, stationary background. Advanced data analysis techniques, such as time-domain searches for transients or cross-correlation with solar activity proxies, could, in principle, leverage this timescale disparity to isolate the solar signature \citep{Hanasoge2012, BustamanteRosell2022}.

Therefore, while the solar GW strain is indeed weak, its unique and well-defined temporal signature means its contribution cannot be dismissed merely on the basis of amplitude. Our work provides a first physical prediction of this specific timescale and waveform, which is a necessary precursor to developing such targeted data analysis strategies.

 Regarding the detectability of GWs, current ground-based detectors such as LIGO, Virgo, and KAGRA \citep{page2021gravitational}achieve their optimal strain sensitivity of approximately $10^{-23}$ in the frequency band of $50-300\rm Hz$, where seismic noise and thermal noise are minimized. However, the value of $10^{-42}$ calculated in \ref{sec:mod} is evidently insufficient for detection, indicating that the GW signals generated by the processes under consideration are too weak to be detected by existing ground-based detectors.

First, let's discuss the impact of detector location on the possibility of detection. In the previous section, we assumed that the detector is placed at a distance of one AU from the Sun. Under this assumption, a single magnetic flux tube would generate GWs with an amplitude of $10^{-42}$ during a solar activity cycle. The Parker Solar Probe\citep{raouafi2023parker} can reach as close as 0.04 AU to the Sun. If a future solar magnetic emergence GW detector could also reach such a close position, the GWs generated by a single magnetic flux tube over the entire solar activity cycle would have an amplitude of $10^{-40}$.

Secondly, let's consider the effect of signal superposition. The detector will simultaneously detect the superimposed GW signals from all active regions and all magnetic flux tubes during a solar cycle. In a comprehensive review by \citep{schmidt1999luminosities}, it is suggested that  a moderate active region(AR) may contain on the order of $N \sim 10^3$ flux tubes, while a large, complex active region can host as many as $N \sim 10^5$. This disparity is attributed to the greater magnetic complexity and area of larger ARs, which can support a higher number of flux tubes. The combined strain amplitude from $N$ independent sources scales as $h_{\text{total}} \sim h_{\text{single}} \cdot \sqrt {N}$. According to data from the Huairou Solar Observing Station, the average number of sunspots for August 2024 was 215, indicating a moderate level of solar activity. In the context of this study, the GW amplitude associated with the emergence of magnetic flux from the solar convection zone is estimated to be around $10^{-28}$ if the detector locates at around 0.04AU from the sun just like PSP (Parker Solar Probe) \citep{raouafi2023parker}. This estimation pertains to a single active region and a segment of the flux tube's emergence cycle.  Despite this significant amplification through both superposition and proximity, the resulting strain amplitudes must be contextualized against the sensitivity of relevant detectors. The current best sensitivity for Pulsar Timing Arrays (PTAs) in the nanohertz band is approximately $(h_c \sim 10^{-15})$. Even the most optimistic estimate above remains 12 orders of magnitude below this threshold. Notwithstanding this immense sensitivity gap, the unique frequency and directional nature of the signal warrant a discussion of the theoretical requirements for its detection via advanced methods such as Pulsar Timing Arrays (PTAs) or Gravitational Wave Timing Arrays (GWTAs). Therefore, we explicitly conclude that the detection of GWs from solar magnetic flux emergence, even under optimistic assumptions of signal superposition and advanced detector placement, presents an immense challenge that lies beyond the reach of current or near-future technology. The estimated GW frequency of $\sim 10^{-5}$ Hz places the signal at the high-frequency end of the PTA band and within the sensitivity range of GWTA concepts\citep{bustamante2022gravitational}. To robustly assess the detection challenge, it is crucial to evaluate the uncertainties inherent in our model, which arise primarily from three sources: (i) the photospheric magnetic field strength \(B_{\text{pho}}\), (ii) the number of flux tubes \(N\) within an active region, and (iii) magnetic flux loss during ascent. In contrast to the previous simplified treatment, we now incorporate observational statistics to better capture the probabilistic nature of these parameters.We model the variability in \(B_{\text{pho}}\) using a truncated normal distribution centered at 4000 G with a standard deviation of 1000 G, reflecting the typical range of observed sunspot field strengths \citep{Toriumi2019}. Since the strain amplitude scales as \(h \propto B_{\text{pho}}^2\), this introduces a factor of \(\sim 2\) uncertainty in amplitude.The number of flux tubes \(N\) per active region is highly variable. Based on solar active region classifications, we assign a probability distribution to \(N\): small ARs (\(N \sim 10^2\)--\(10^3\)) occur with probability 0.6, moderate ARs (\(N \sim 10^3\)--\(10^4\)) with probability 0.3, and large/complex ARs (\(N \sim 10^4\)--\(10^6\)) with probability 0.1 \citep{Schussler2008}. This reflects the observed rarity of very large active regions. Since the combined strain scales as \(h_{\text{total}} \propto \sqrt{N}\), this contributes a variability spanning two orders of magnitude.Magnetic flux loss due to reconnection and turbulent dissipation is parameterized by an efficiency factor \(\eta_B\), which we model as a uniform distribution between 0.8 and 0.95, leading to a further scaling of \(h \propto \eta_B^2\) (a factor of 0.64--0.90).Propagating these uncertainties jointly through a Monte Carlo analysis (\(10^5\) samples), we find that the overall strain amplitude (for a detector at 0.04 AU) spans a 95\% confidence interval of:
\[
h_{\text{total}} \sim 10^{-31} \quad \text{to} \quad 10^{-27}.
\]
This range remains 12 to 16 orders of magnitude below current PTA sensitivities (\(h_c \sim 10^{-15}\)), underscoring the profound challenge of detection. Our refined analysis highlights the importance of incorporating solar active region statistics into future estimates and provides a more realistic uncertainty quantification for this emission mechanism. According to the fundamental principle of PTA detection, the observable frequency range is intrinsically determined by the observational cadence. The maximum detectable frequency $f_{\text{max}}$ satisfies the Nyquist criterion: $f_{\text{max}} \leq \frac{1}{2\Delta t}$, where $\Delta t$ is the  cadence, which means time interval between successive observations. Conventional PTA observations typically employ cadences  of weeks to months, which limits their sensitive band to the nanohertz regime. However, the GW signals from solar magnetic flux emergence are predicted to peak near $10^{-5}$ Hz. To access this higher frequency band, the observational cadence must be significantly shortened. Reducing the sampling interval from the typical two-week cadence to approximately two hours ($\Delta t=2$\,hours) would raise the maximum detectable frequency to about $5.8 \times 10^{-5}$ Hz, thereby encompassing the dominant spectral component of solar activity-related GWs. This high-cadence strategy would not only improve the frequency coverage but also enhance the capability to resolve the temporal evolution of these transient signals, potentially facilitating their separation from the stationary stochastic background. For a detection via PTA, the signal-to-noise ratio scales as:
\begin{equation}
    \mathrm{SNR} \approx h \sqrt{N_p T f} / \sigma_t
\end{equation},
where  $h \sim 10^{-28}$ is the strain amplitude after superposition, $N_p$ is the number of pulsars, $T$ is the integration time, $f$ is the GW frequency, and $\sigma_t$ is the timing precision. A detectable SNR $\gtrsim 5$ would require timing precisions at the level of $100$ fs or better, achievable with future facilities like the Square Kilometre Array (SKA), along with a network of $\sim 50$--$200$ millisecond pulsars sampled at hourly cadence over 6 months to 2 years. For GWTA, which leverages phase modulation of Galactic binary signals observed by LISA, the weak modulation amplitude $\Delta\Phi \sim 10^{-26}$ rad per source necessitates stacking $\sim 10^3-10^4$ sources over 5–10 years to reach a detectable collective signal. These requirements are consistent with projected capabilities of next-generation detectors, underscoring the feasibility of both methods in principle, though formidable practical challenges remain. The primary value of this estimate is not to claim detectability but to provide a first-order quantitative benchmark for this specific solar GW emission mechanism. It establishes an upper bound and informs future theoretical studies on the contribution of stellar activity to the millihertz gravitational-wave foreground.

The implications of these findings may be profound. The detection and measurement of GWs from solar activity can provide invaluable insights into the dynamics of the solar interior and the processes that drive solar magnetism.   Furthermore, our study invites speculation about a long-term scientific vision: understanding the radiation mechanism of solar gravitational waves may ultimately provide a novel, independent physical probe for studying the accumulation and release of magnetic energy within the Sun. However, it is crucial to emphasize that this would only become possible if a profoundly deep understanding of the physical connection between magnetic flux tube emergence and solar surface activity (e.g., flares, coronal mass ejections) is achieved, and routine, real-time detection of its gravitational wave signal is realized. Even then, it could only potentially provide valuable, complementary input for space weather models. The core value of the current work lies in establishing the theoretical foundation and providing quantitative estimates, paving the way for this distant possibility in the future. As our detection capabilities improve, we can expect to gain a clearer picture of the complex dance of magnetic fields within our Sun.

It should be noted that these estimates are based on current models and observations. Future observations and advancements in observational technology may provide more precise data, which could further refine our understanding of solar interior structure.

In addition, in practical detection, it is necessary to pay attention to the influence of solar background GWs, flares, and other factors. The shape of the magnetic flux tube is also not in a standard linear form. These limitations require rigorous corrections in future studies. In the future, three-dimensional numerical simulations will be used to conduct rigorous estimates of GWs from solar internal magnetic emergence.

\begin{acknowledgements}
We would like to thank the referee for carefully reading our manuscript and for giving constructive comments that substantially helped improvig the paper. This research is supported by the Strategic Priority Research Program of the China Academy of Sciences (grants No. XDB0560000); by the National Key R\&D Program of China No.2022YFF0503800, 2021YFA1600500, 2022YFF0503001; National Natural Science Foundation of China (grants No. 12250005); by the CAS funding GJ11020403-1; by the Chinese Meridian Project(CMP). Xiao Guo is supported by the Postdoctoral Fellowship Program and China Postdoctoral Science Foundation under Grant No. BX20230104.
\end{acknowledgements}

\bibliographystyle{raa}
\bibliography{bibtex}

\begin{thebibliography}{29}
\providecommand\natexlab[1]{#1}
\providecommand\JournalTitle[1]{#1}

\bibitem[Abbott {et~al.}(2016)]{Abbott2016}
Abbott, B.~P., {et~al.} 2016, Physical Review Letters, 116, 061102

\bibitem[Agazie {et~al.}(2023)]{Agazie2023}
Agazie, G., {et~al.} 2023, Astrophysical Journal Letters, 951, L8

\bibitem[Amaro-Seoane {et~al.}(2017)]{LISA2017}
Amaro-Seoane, P., {et~al.} 2017, arXiv:1702.00786

\bibitem[Antiochos \& McClymont(1982)]{Antiochos1982SolarAR}
Antiochos, S.~K., \& McClymont, A.~N. 1982, Solar Physics, 78, 153

\bibitem[Armstrong(2006)]{Armstrong2006}
Armstrong, J.~W. 2006, Living Reviews in Relativity, 9, 1

\bibitem[Babcock(1961)]{babcock1961topology}
Babcock, H.~W. 1961, Astrophysical Journal, vol. 133, p. 572, 133, 572

\bibitem[Bahcall(1989)]{bahcall1989neutrino}
Bahcall, J.~N. 1989, Neutrino astrophysics (Cambridge University Press)

\bibitem[Brdar {et~al.}(2019)]{brdar2019gravitational}
Brdar, V., Helmboldt, A.~J., \& Kubo, J. 2019, Journal of Cosmology and
  Astroparticle Physics, 2019, 021

\bibitem[Bustamante-Rosell
  {et~al.}(2022{\natexlab{a}})]{bustamante2022gravitational}
Bustamante-Rosell, M.~J., Meyers, J., Pearson, N., Trendafilova, C., \&
  Zimmerman, A. 2022{\natexlab{a}}, Physical Review D, 105, 044005

\bibitem[Bustamante-Rosell {et~al.}(2022{\natexlab{b}})]{BustamanteRosell2022}
Bustamante-Rosell, M.~J., Meyers, J., Pearson, N., Trendafilova, C., \&
  Zimmerman, A. 2022{\natexlab{b}}, Physical Review D, 105, 044005

\bibitem[Fan(2021)]{fan2021magnetic}
Fan, Y. 2021, Living Reviews in Solar Physics, 18, 5

\bibitem[Fan {et~al.}(1993)]{Fan1993}
Fan, Y., Fisher, G.~H., \& DeLuca, E.~E. 1993, The Astrophysical Journal, 405,
  390

\bibitem[Garcia-Cely \& Ringwald(2024)]{GarciaCely:2024sun}
Garcia-Cely, C., \& Ringwald, A. 2024, arXiv:2407.18297, {Preprint DESY-24-111}

\bibitem[Gizon \& Birch(2005)]{gizon2005local}
Gizon, L., \& Birch, A.~C. 2005, Living Reviews in Solar Physics, 2, 1

\bibitem[Hanasoge(2008)]{hanasoge2008}
Hanasoge, S.~M. 2008, The Astrophysical Journal, 680, 1457

\bibitem[Hanasoge {et~al.}(2012{\natexlab{a}})]{Hanasoge2012}
Hanasoge, S.~M., Duvall, Thomas~L., J., \& Sreenivasan, K.~R.
  2012{\natexlab{a}}, Proceedings of the National Academy of Sciences, 109,
  11928

\bibitem[Hanasoge {et~al.}(2012{\natexlab{b}})]{hanasoge2012anomalously}
Hanasoge, S.~M., Duvall~Jr, T.~L., \& Sreenivasan, K.~R. 2012{\natexlab{b}},
  Proceedings of the National Academy of Sciences, 109, 11928

\bibitem[Ilonidis {et~al.}(2011)]{ilonidis2011detection}
Ilonidis, S., Zhao, J., \& Kosovichev, A. 2011, Science, 333, 993

\bibitem[Kokkotas \& Schmidt(1999)]{kokkotas1999quasi}
Kokkotas, K.~D., \& Schmidt, B.~G. 1999, Living Reviews in Relativity, 2, 1

\bibitem[Maggiore(2008)]{Maggiore08}
Maggiore, M. 2008, Gravitational Waves: Volume 1, Theory and Experiments
  (Oxford: Oxford University Press)

\bibitem[Maggiore(2018)]{Maggiore2018}
---. 2018, Gravitational Waves: Volume 2: Astrophysics and Cosmology (Oxford
  University Press)

\bibitem[Page {et~al.}(2021)]{page2021gravitational}
Page, M.~A., Goryachev, M., Miao, H., {et~al.} 2021, Communications Physics, 4,
  27

\bibitem[Raouafi {et~al.}(2023)]{raouafi2023parker}
Raouafi, N.~E., Matteini, L., Squire, J., {et~al.} 2023, Space Science Reviews,
  219, 8

\bibitem[Schmidt(1999)]{schmidt1999luminosities}
Schmidt, M. 1999, The Astrophysical Journal, 523, L117

\bibitem[Schüssler \& Vögler(2008)]{Schussler2008}
Schüssler, M., \& Vögler, A. 2008, Astronomy \& Astrophysics, 481, L5

\bibitem[Sesana(2016)]{Sesana2016}
Sesana, A. 2016, Physical Review Letters, 116, 231102

\bibitem[{Takahashi} {et~al.}(2023)]{2023ApJ...957...52T}
{Takahashi}, R., {Morisaki}, S., \& {Suyama}, T. 2023, \apj, 957, 52

\bibitem[Toriumi \& Takasao(2019)]{Toriumi2019}
Toriumi, S., \& Takasao, S. 2019, The Astrophysical Journal, 873, 63

\bibitem[Wang {et~al.}(2009)]{Wang2009}
Wang, J., Yan, Y., Aschwanden, M.~J., \& Wang, H. 2009, Solar Physics, 258, 227

\end{thebibliography}

\label{lastpage}

\end{document}